%% file: main.tex
 \pdfoutput=1
\documentclass[sigconf,nonacm]{acmart}
\AtBeginDocument{%
  \providecommand\BibTeX{{%
    \normalfont B\kern-0.5em{\scshape i\kern-0.25em b}\kern-0.8em\TeX}}}

\usepackage{algorithm}
\usepackage{algpseudocode}

\input{packages}

\begin{document}

\title{Fractional Budget Allocation for Influence Maximization under General Marketing Strategies}

% \author{Akhil Bhimaraju, Eliot W. Robson, Lav P. Varshney, and Abhishek K. Umrawal}
% \email{{akhilb3, erobson2, varshey, aumrawal}@illinois.edu}
% \affiliation{%
%   \institution{University of Illinois Urbana Champaign}
%   \city{Urbana}
%   \state{IL}
%   \postcode{61801-2332}
% }

\author{Akhil Bhimaraju}
\email{akhilb3@illinois.edu}
\affiliation{%
  \institution{University of Illinois Urbana Champaign}
  \city{Urbana}
  \state{IL}
  \country{USA}
  \postcode{61801-2332}
}

\author{Eliot W. Robson}
\email{erobson2@illinois.edu}
\affiliation{%
  \institution{University of Illinois Urbana Champaign}
  \city{Urbana}
  \state{IL}
  \country{USA}
  \postcode{61801-2332}
}

\author{Lav R. Varshney}
\email{varshney@illinois.edu}
\affiliation{%
  \institution{University of Illinois Urbana Champaign}
  \city{Urbana}
  \state{IL}
  \country{USA}
  \postcode{61801-2332}
}

\author{Abhishek K. Umrawal}
\email{aumrawal@illinois.edu}
\affiliation{%
  \institution{University of Illinois Urbana Champaign}
  \city{Urbana}
  \state{IL}
  \country{USA}
  \postcode{61801-2332}
}

\input{sections/abstract}





\maketitle
\pagestyle{empty}

\input{sections/introduction}
\input{sections/model}
\input{sections/algorithm}
\input{sections/experiments}
\input{sections/conclusion}

% \begin{acks}
% \end{acks}

%\clearpage
\bibliographystyle{ACM-Reference-Format}
\bibliography{refs}

% \appendix









\end{document}

%% file: packages.tex
\usepackage{amsmath}
\usepackage{amsthm}
\newtheorem{theorem}{Theorem}

\usepackage{algorithm}
\usepackage{algpseudocode}
\usepackage{multirow}
\usepackage{multicol}
\usepackage{mathtools}

\def\BibTeX{{\rm B\kern-.05em{\sc i\kern-.025em b}\kern-.08em
    T\kern-.1667em\lower.7ex\hbox{E}\kern-.125emX}}

\usepackage[utf8]{inputenc}
\usepackage{pgfplots}
\usepgfplotslibrary{groupplots,dateplot}
\usetikzlibrary{patterns,shapes.arrows}
\pgfplotsset{compat=newest}
\pgfplotsset{scaled y ticks=false}
\usepackage{caption}
\usepackage{subcaption}
\usepackage{caption}
\usepackage{subcaption}
\usepackage{amsmath}
\usepackage{mathtools}

\usepackage{algorithm}
\usepackage{algpseudocode}
\usepackage{xcolor}
\usepackage{soul}

\usepackage[capitalize,noabbrev]{cleveref}
\crefname{section}{Section}{sections}
\crefname{problem}{Problem}{problems}

\usepackage[utf8]{inputenc}
\usepackage{pgfplots}

\usepgfplotslibrary{groupplots,dateplot}
\usetikzlibrary{patterns,shapes.arrows}
\pgfplotsset{compat=newest}

\usepackage{comment}

\includecomment{commentCDCtrim}
\excludecomment{commentCDCtrim}

\includecomment{addclearpage}
\excludecomment{addclearpage}

%% file: sections/abstract.tex
\begin{abstract}
  % We consider the problem of maximizing influence over a social network by giving discounts
  % to some nodes in the network.
  % A greater discount to a particular node increases the probability of its activation, which in turn
  % tries to activate its neighbors leading to the influence process cascading through the network.
  % The objective is to develop fast algorithms that give us initial discounts to the various
  % nodes in the network to maximize the total number of nodes activated by the process, subject to a
  % sum constraint on the total discount given.
  % While recent work has given an algorithm for this problem when the relation between the discount
  % given and the activation probability is the identity function, we extend it to the case where the
  % activation probabilities are general affine functions of the discount.
  % Since the problem is NP-hard to solve optimally, the performance guarantees are in terms of
  % approximation factors, and we show our algorithm has the same (constant-factor) guarantee
  % as prior work.
  % We also show that our algorithm is much faster than comparable methods via
  % numerical experiments on real-world graphs.
    We consider the \textit{fractional} influence maximization problem, i.e., identifying users on a social network to be incentivized with potentially \textit{partial} discounts to maximize the influence on the network. The larger the discount given to a user, the higher the likelihood of its activation (adopting a new product or innovation), who then attempts to activate its neighboring users, causing a cascade effect of influence through the network. Our goal is to devise efficient algorithms that assign initial discounts to the network’s users to maximize the total number of activated users at the end of the cascade, subject to a constraint on the total sum of discounts given. In general, the activation likelihood could be any non-decreasing function of the discount, whereas, our focus lies on the case when the activation likelihood is an affine function of the discount, potentially varying across different users. As this problem is shown to be NP-hard, we propose and analyze an efficient $(1-1/e)$-approximation algorithm. Furthermore, we run experiments on real-world social networks to show the performance and scalability of our method. % in contrast to other methods.
\end{abstract}

%% file: sections/introduction.tex
\section{Introduction}
\label{sec:introduction}

Marketing firms who wish to promote their products or services on a social network should solve the problem of selecting influential nodes on the network who can spread the word among their followers \cite{KempeKT2003,umrawal2023community,chen2020scalable}. Once the influencers have been identified, the firm offers discounts to incentivize them to use and promote the product in their network. The higher the discount given to a particular influencer, the more likely they will use the product and promote it among their followers. Since marketing budgets tend to be finite, the total amount of discount given is limited by a sum constraint. In the binary-discount case where a node is either completely selected or not (i.e., the ``discount'' given is 100\%), this problem reduces to a set-selection problem under a cardinality constraint where we need to select the optimal set of nodes as our influencers. This problem has been modeled using standard diffusion models over the network. In their seminal work, Kempe et al. \cite{KempeKT2003} showed that this set-selection problem involves solving a submodular maximization problem subject to a cardinality constraint. They also show that this problem is NP-hard, and using techniques from \cite{nemhauser1978analysis}, they develop greedy algorithms and show that the solution given by their algorithm is at least $\left(1-1/e\right)$ times the optimal solution. Following \cite{kempe2003maximizing}, several solutions \cite{leskovec2007cost,goyal2011celf++,goyal2011data,umrawal2023leveraging,umrawal2023community} have been proposed this problem.

For the continuous case where discounts are arbitrary, \cite{KempeKT2003} also provides a hill-climbing algorithm that traverses the feasible set in steps similar to a gradient ascent. The submodular maximization problem has seen a lot of significant work over the years, and we now have algorithms for more general constraints \cite{Vondrak2008,CalinescuCPV2011}. However, these still study the discrete optimization problem where the discounts are binary. More recently, a much faster algorithm for the continuous problem where potentially fractional discounts are given to the influencers has been developed by \cite{UmrawalAQ2023}. They achieve this by hopping over vertices of the feasible set directly and avoiding the more time-consuming hill-climbing process. However, a key assumption in their setup is that the discount given is equal to the probability of a node getting activated, i.e., the probability that the influencer uses the product and advertises it in their network. This is a very restrictive assumption since there is no reason ``discount'' and ``probability of activation'' should be the same. It is not even clear if they are in the same space in terms of units and dimensions. In this paper, we extend their work to the more general case where the probability of activation is any arbitrary affine function of the discount offered, with potentially different functions for different nodes. We prove that our algorithm achieves the same $\left(1-1/e\right)$ approximation guarantee as in prior work, and demonstrate its effectiveness in numerical simulations over real-world networks.

The rest of this paper is organized as follows. Sec.~\ref{sec:model} describes our notation, the mathematical model, and the formal problem. Sec.~\ref{sec:algorithm} discusses our algorithm and proves the constant-factor approximation guarantee. Sec.~\ref{sec:experiments} shows the empirical performance of our algorithm on real-world networks, and Sec.~\ref{sec:conclusion} concludes the paper.

%% file: sections/model.tex
\section{Mathematical Model}
\label{sec:model}
We mostly follow the notation used in \cite{UmrawalAQ2023}. Consider a social network $G=(V,E)$ where a node $v\in V$ represents a person who can influence their neighbors $\mathcal{N}_v=\{u:(v,u)\in E\}$ in the edge set $E$. At time $0$, node $v$ is given a \emph{discount} $y_v$,  which \emph{activates} $v$ with probability $f_v(y_v)$. Let the initially active set be denoted by $S_0$. At each time $t$, if an active node $v$ is the neighbor of an inactive node $u$, the node $u$ gets activated with probability $e_{v,u}$ independently for each active neighbor $v$. More concretely, the probability that $u$ gets activated at time $t$, $p_u^{(t)}$ is given by
\begin{align*}
p_u^{(t)}=1-\prod_{\{v:u\in\mathcal{N}_v\}\cap S_{t-1}}\left(1-e_{v,u}\right),
\end{align*}
where $S_t$ denotes the set of nodes activated at time $t$ (which does not include the nodes activated before time $t$). Note that in this model, nodes do not get deactivated, and each activated node has only one attempt at activating each of its inactive neighbors. The goal is to find the optimal discounts $y_v$ to be given under a sum constraint $\sum_{v\in V}y_v \le K$, so that the most number of nodes are activated at the end of the cascade (in expectation):
\begin{align}
    F(y) = \mathbb{E}\left[\text{card}\left(\bigcup_{t=0}^\infty S_t\right) \mid y\right],
    \label{eq:objective-infty}
\end{align}
where $y$ denotes the vector containing the values of $y_v$, and $\text{card}(\cdot)$ denotes cardinality\footnote{While we use $S_\infty$ for notational convenience, note that the cascade terminates after a finite number of time steps.}. The optimization problem for \textit{generalized} fractional influence maximization \eqref{eq:cont-optimization} is formally stated as follows.
\begin{align}
    y^* = \arg&\max_{y_v\ge0} F(y) \tag{GFIM}\label{eq:cont-optimization}\\
    \text{subject to}&\ \sum_{v\in V} y_v \le K,\nonumber\\
    \text{and}\ \ &\ \ f_v(y_v) \le 1\ \ \text{for all} \ v\in V.\nonumber
\end{align}
%In their landmark paper, 
\cite{KempeKT2003} show that the final objective can be expressed as a submodular function of $S_0$, $\sigma(S_0)$.
This implies we can rewrite \eqref{eq:objective-infty} as
\begin{align}
    F(y) &= \mathbb{E}_{S_0}\left[\sigma(S_0) \mid y\right] \nonumber\\
    &= \sum_{S_0\subseteq V} \sigma(S_0)\prod_{v\in S_0}f_v(y_v)\prod_{v\notin S_0}(1-f_v(y_v)),
    \label{eq:objective-submodular}
\end{align}
where $\sigma(\cdot)$ is a submodular function on $V$, i.e.,
\begin{align}
    \sigma(R\cup S) + \sigma(R\cap S) \le \sigma(R) + \sigma(S) \quad \forall\ R,S\subseteq V.
    \label{eq:submodulat-definition}
\end{align}

An approximation algorithm referred to as MLE-Greedy was provided in \cite{UmrawalAQ2023} to solve \eqref{eq:cont-optimization} when $f_v(y)=y$ for all $y\in[0,1]$ and $v\in V$. They referred to that simpler case as fractional influence maximization (FIM). This allows \eqref{eq:objective-submodular} to be interpreted as a multilinear extension from \cite{Vondrak2008}.
However, this is very restrictive and does not capture the way nodes get activated in the real world. There is no reason the amount of discount should be equal to the probability of the node getting activated\footnote{It is not even clear if the ``discount'' and ``probability of activation'' are in the same space in terms of units, dimensions, etc.}. In this work, we directly analyze \eqref{eq:objective-submodular} and develop a polynomial-time constant-factor approximation algorithm when $f_v(\cdot)$ are any increasing linear functions.

%% file: sections/algorithm.tex
\section{Approximation algorithm with constant-factor guarantee}
\label{sec:algorithm}

In this section, we provide our algorithm for \eqref{eq:cont-optimization} and prove its performance guarantees. The algorithm is based on greedy optimization, which has long been used for maximizing submodular functions under a cardinality constraint. The idea is that we start with an empty set and add elements that give the maximum incremental gain in the objective until the constraint is reached and no more elements can be added. A generalization for real-valued submodular functions has been studied by \cite{Wolsey1982} when the objective $F(\cdot)$ is piecewise linear\footnote{Note that this is not true in our case since $F(\cdot)$ is a product of linear functions.}.
A more general problem of submodular maximization under arbitrary matroid constraints has also been studied by \cite{Vondrak2008,CalinescuCPV2011}, but their algorithm involves a continuous-greedy approach that traverses the entire feasible region in small steps. In contrast, the discrete-greedy approach we propose is much faster as its complexity is just polynomial in $|V|$ and does not depend on the problem instance in any other way. We also see in Sec.~\ref{sec:experiments} that our technique is much faster in practice for influence maximization than other techniques that use a continuous-greedy approach. A similar approach has been proposed by \cite{UmrawalAQ2023} but for the special case of $f_v(y)=y$, and the analysis does not seem to directly translate to the more general case. To the best of our knowledge, no prior work uses the discrete approach for more general $f_v(\cdot)$ for the influence maximization problem.

One main difference with the discrete-greedy approach of \cite{UmrawalAQ2023} is that we now also have potentially
different $f_v(\cdot)$ for different $v$.
To account for this, we need to weigh the increment by $f'_v(0)$ before choosing the next best element at each step.
We state the algorithm formally as Alg.~\ref{alg:discrete-greedy-general}.
As in other works on influence maximization, we assume the submodular function $\sigma(\cdot)$
can be computed.
In practice, this is usually estimated using simulations.

\begin{algorithm}
    \caption{Generalized discrete-greedy algorithm for \eqref{eq:cont-optimization}}
    \label{alg:discrete-greedy-general}
    \textbf{Input:} Nodes $V$, Submodular function $\sigma:2^V\mapsto\mathbb{R}_{\ge0}$, max. sum $K$\\
    \textbf{Output:} Vector $y\in\mathbb{R}_{\ge0}^{|V|}$, where $y_v$ denotes discount to node $v$\\
    \begin{algorithmic}[1]
        \State Initialize $y_v\gets0$ for all $v\in V$ and $S_0\gets\emptyset$
        \While {$\sum_{v\in V}y_v < K$ \textbf{and} $S_0\neq V$}
            \State $v_\text{next}\gets\arg\max_{v\notin S_0} (\sigma(S_0\cup \{v\})-\sigma(S_0))f'_v(0)$
            \label{alg:step:choose-max}
            \State $y_{v_\text{next}}\gets\min\left(f_{v_\text{next}}^{-1}(1), K-\sum_{v\in V}y_v\right)$
            \label{alg:step:vnext}
            \State $S_0\gets S_0\cup\{v_\text{next}\}$
        \EndWhile
    \end{algorithmic}
\end{algorithm}

We state the performance guarantee of Alg.~\ref{alg:discrete-greedy-general} as Theorem~\ref{thm:const-approx}.
\begin{theorem}
When the functions $f_v(\cdot)$ are all nonnegative increasing linear functions,
i.e., $f_v(y_v)=a_vy_v+b_v$ with $a_v,b_v\ge0$, the output $y$ of Alg.~\ref{alg:discrete-greedy-general} satisfies
\begin{align*}
    F(y) \ge F(y^*)\left(1-\frac{1}{e}\right),
\end{align*}
where $y^*$ is the optimal solution of \eqref{eq:cont-optimization}.
Further, Alg.~\ref{alg:discrete-greedy-general} runs in $O\left(|V|^2T\right)$, where $T$ is the time complexity of
computing $\sigma(\cdot)$.
\label{thm:const-approx}
\end{theorem}
\begin{proof}
    The proof is based on techniques from \cite{Vondrak2008,CalinescuCPV2011}.
    Let the output of Alg.~\ref{alg:discrete-greedy-general} for a given value of $K$ be
    denoted by $y^{(K)}$,
    and let the set of nodes whose probabilities of getting selected are $1$ for that value of
    $K$ be given by $S_0^{(K)}$, i.e., $S_0^{(K)}=\{v:f_v(y_v^{(K)})=1\}$.
    Then we can write
    \begin{align}
        \frac{d}{dK}F(y^{(K)}) = \sum_{v\in V}\frac{\partial F(y^{(K)})}{\partial y_v}\frac{dy^{(K)}_v}{dK}.
        \label{eq:FK-chain-rule}
    \end{align}

    Differentiating \eqref{eq:objective-submodular}, we get
    \begin{align}
        \frac{\partial F(y)}{\partial y_v}\!\! = \!\!\!\!\!\!\sum_{S_0\subseteq V\setminus\{v\}}\!\!\!\!\!\!\!\!\!\left(\sigma(S_0\cup\{v\})\!-\!\sigma(S_0)\right)f'_v(y_v)
        \!\!\prod_{u\in S_0}\!\!f_u(y_u)\!\!\prod_{u\notin S_0}\!\!(1\!-\!f_u(y_u)).
        \label{eq:first-derivative}
    \end{align}
    Observe that in Alg.~\ref{alg:discrete-greedy-general}, at most one node can have a probability of getting selected as neither
    $0$ nor $1$ (the last node selected), and let us denote this node by $v_{\text{last}}^{(K)}$.
    Since $f_u(y_u)$ is either $0$ or $1$ depending on whether $u$ is in $S_0^{(K)}$ or not, for the values of $y^{(K)}$ output
    by Alg.~\ref{alg:discrete-greedy-general}, the partial derivative reduces to
    \begin{align}
        \frac{\partial F(y^{(K)})}{\partial y_{v_{\text{last}}}}=\left(\sigma(S_0^{(K)}\cup\{v_{\text{last}}^{(K)}\})-\sigma(S_0^{(K)})\right)f'_{v_{\text{last}}}(y_{v_{\text{last}}}^{(K)}).
        \label{eq:partial-simplification}
    \end{align}
    Moreover, for $v\in S_0^{(K)}$, we have $v\neq v_{\text{last}}^{(K)}$ and $S_0^{(K)}\cup\{v\}=S_0^{(K)}$, which implies
    \begin{align}
        \frac{\partial F(y^{(K)})}{\partial y_{v}}\!=\!\!
        \begin{cases}
            \left(\sigma(S_0^{(K)}\cup\{v\})\!-\!\sigma(S_0^{(K)})\right)f'_{v}(0) \ \text{if}\ v\notin S_0^{(K)}\!\!,v\neq v_{\text{last}}^{(K)},\\
            \qquad\qquad\qquad 0 \qquad\qquad \text{otherwise}.
        \end{cases}
        \label{eq:partial-simplification-general}
    \end{align}
    Further, as $K$ increases infinitesimally, the only node whose probability is affected is $v_{\text{last}}^{(K)}$,
    and from step~\ref{alg:step:vnext} of Alg.~\ref{alg:discrete-greedy-general}, we can see that the derivative with respect
    to $K$ is $1$.
    Substituting this observation and \eqref{eq:partial-simplification} into \eqref{eq:FK-chain-rule}, we get
    \begin{align}
        \frac{d}{dK}F(y^{(K)}) = \left(\sigma(S_0^{(K)}\cup\{v_{\text{last}}^{(K)}\})-\sigma(S_0^{(K)})\right)f'_{v_{\text{last}}}(y_{v_{\text{last}}}^{(K)}).
        \label{eq:derivative-wrt-K}
    \end{align}
    
    Let $y^*$ denote the optimal value of \eqref{eq:cont-optimization} for a given value of $K=C$.
    We wish to analyze the behavior of \eqref{eq:derivative-wrt-K} as the value of $K$ increases from
    $0$ to $C$.
    Consider the function $F\left(y+\xi(y^*-y)^+\right)$ for $\xi\in[0,1]$.
    Differentiating \eqref{eq:first-derivative} and using the fact that $f''_v(y)=0$ since
    $f_v(\cdot)$ is linear, we get $\frac{\partial^2F}{\partial y_v\partial y_u}\le0$
    for all $u,v\in V$ (with equality for $u=v$).
    This implies that the maximum value of $\frac{dF(y+\xi(y^*-y)^+)}{d\xi}$ is at $\xi=0$ since $(y^*-y)^+\ge0$.
    This gives for all $\xi\in[0,1]$
    \begin{align*}
        \frac{d F\left(y^{(K)}+\xi(y^*-y^{(K)})^+\right)}{d\xi}&\le \left.\frac{d F\left(y^{(K)}+\xi(y^*-y^{(K)})^+\right)}{d\xi}\right|_{\xi=0}\\
        &=\sum_{v\in V}\frac{\partial F(y^{(K)})}{\partial y_v}(y^*-y^{(K)})^+_v \\
        &\le \max_{v\in V}\left\{\frac{\partial F(y^{(K)})}{\partial y_v}\right\}\!\!\sum_{v\in V}\!\!\left(y^*-y^{(K)}\!\right)^+_v\!\!.
    \end{align*}
    Since $\frac{\partial^2F}{\partial y_v\partial y_u}\le0$ for $u\neq v$ and $\frac{\partial^2F}{\partial y_v^2}=0$,
    step~\ref{alg:step:choose-max} ensures that the $\max$ term in the inequality always occues at $v=v_{\text{last}}$.
    Further, $y^*$ must be a valid solution, which gives $\sum_{v\in V}(y^*-y)^+\le C$ (recall that $y^*$ is the solution
    for a given $K=C$).
    Together with \eqref{eq:partial-simplification} and \eqref{eq:derivative-wrt-K}, these observations give for $\xi\in[0,1]$,
    \begin{align}
        \frac{d F\left(y^{(K)}+\xi(y^*-y^{(K)})^+\right)}{d\xi}&\le C\frac{d}{dK}F(y^{(K)}).
        \label{eq:general-xi-le-K}
    \end{align}

    We also have
    \begin{align}
        F(y^*) - F(y^{(K)}) &\overset{\text{(a)}}{\le} F\left(y^{(K)}+(y^*-y^{(K)})^+\right) - F(y^{(K)})\nonumber \\
        &\overset{\text{(b)}}{=} \left.\frac{d F\left(y^{(K)}+\xi(y^*-y^{(K)})^+\right)}{d\xi}\right|_{\xi=c}
        \label{eq:mean-value}
    \end{align}
    for some $c\in[0,1]$, where $\text{(a)}$ follows from $\frac{\partial F}{\partial y_v}\ge0$ (see \eqref{eq:first-derivative}),
    and $\text{(b)}$ follows from the mean value theorem.
    \eqref{eq:general-xi-le-K} and \eqref{eq:mean-value} together imply    \begin{align*}
        F(y^*) - F(y^{(K)}) \le C\frac{d}{dK}F(y^{(K)}).
    \end{align*}
    So the final output of Alg.~\ref{alg:discrete-greedy-general}, $y^{(C)}$
    dominates the solution of $F(y^*)-\phi=C\frac{d\phi}{dK}$ at $K=C$.
    This gives
    \begin{align*}
        F(y^{(C)}) \ge \left(1-\frac{1}{e}\right)F(y^*),
    \end{align*}
    which concludes the proof.
\end{proof}

%% file: sections/experiments.tex
\section{Experiments}
\label{sec:experiments}

In this section, we provide experiments where we evaluate our greedy approximation algorithm on large real-world networks.

\textbf{Network Data.} We ran experiments using three real-world social network graphs from \cite{snapnets}. The number of nodes, number of edges, and type of each network are provided in \cref{tab:basic_info}. 
% Facebook \cite{leskovec2012learning} network is a graph representing circles (or friends lists) from Facebook. Wikipedia \cite{leskovec2010signed,leskovec2010predicting} network is a who-votes-on-whom graph of Wikipedia users to become an administrator. Deezer \cite{feather} network is a graph of mutual follower relationships among Deezer users from Europe. 
Each edge in the undirected networks is replaced by two directed edges. For edge weights, we use the \textit{weighted cascade model} \cite{kempe2003maximizing} where for each node $v\in V$, the weight of each edge entering $v$ was set to $1/\text{in-degree}(v)$. We used the Pooch library \cite{Uieda2020} for automated data retrieval and the NetworkX library \cite{networkx} for graph data processing.

% \textbf{Network Data.} We ran experiments using three real-world social network graphs from the {Stanford Large Network Dataset Collection} \cite{snapnets}. The number of nodes, number of edges, and type of each network are provided in \cref{tab:basic_info}. Facebook \cite{leskovec2012learning} network is a graph representing circles (or friends lists) from Facebook. Wikipedia \cite{leskovec2010signed,leskovec2010predicting} network is a who-votes-on-whom graph of Wikipedia users to become an administrator. %Facebook-Page \cite{rozemberczki2019multiscale} network is a page-page graph of verified Facebook sites/pages. 
% Deezer \cite{feather} network is a graph of mutual follower relationships among Deezer users from Europe. Each edge in the undirected networks is replaced by two directed edges. For edge weights, we use the \textit{weighted cascade model} \cite{kempe2003maximizing} where for each node $v\in V$, the weight of each edge entering $v$ was set to $1/\text{in-degree}(v)$. We also used the Pooch library \cite{Uieda2020} for automated data retrieval and the NetworkX library \cite{networkx} for graph data processing.

\begin{table}[ht]
\caption{Basic information of the networks used} \label{tab:basic_info}
\centering
\begin{tabular}{c c c c}
    \hline \hline
    Network & Nodes & Edges & Type\\
    \hline
    Facebook \cite{leskovec2012learning} & 4,039 & 88,234 & Undirected \\
    Wikipedia  \cite{leskovec2010signed,leskovec2010predicting} & 7,115 & 103,689 & Directed \\    
    Deezer \cite{feather} & 28,281 & 92,752 & Undirected \\
    \hline
\end{tabular}
\end{table}

\textbf{Algorithms.} We run experiments with the proposed generalized discrete-greedy algorithm using non-negative linear functions as in Theorem \ref{thm:const-approx}. We do this by running the algorithm with different sets of random non-negative coefficients for the linear functions $f_v$ at each vertex $v$. As a baseline, we include runs where the coefficients are fixed to $a_v = 1$ and $b_v = 0$, which in the case of an integral value of $k$, recovers the behavior of the Floor-Greedy algorithm (using the standard weighting scheme and ignoring fractional weights).

%As the proposed MLE-Greedy algorithm employs a subroutine for solving DIM, we use the state-of-the-art RIS-based greedy algorithm \cite{guo2020influence} as a black-box DIM solver subroutine. To assess the performance of the proposed greedy algorithm, we use the following baselines.
%\begin{enumerate}
%    \item Floor-Greedy -- for any budget $k$, calculate \textsc{Oracle}($F,\floor{k}$), i.e. ignore the fractional budget.
%    \item LIM-Greedy \cite{chen2020scalable} -- a state-of-the-art, RIS-based adaptation of gradient methods for DR-submodular maximization for the problem of IM.
%\end{enumerate}

\textbf{Experimental Details.} %For LIM-Greedy \cite{chen2020scalable}, we set the discretization parameter, $\delta = 0.1$ and approximation parameter, $\epsilon=0.5$ as suggested by the authors. For the RIS-based greedy, we set the approximation parameter, $\epsilon=0.01$ as suggested by the authors. For our experiments, we compared our method and the baselines in terms of the influence of the output partial incentive vector and empirical run times. 
We ran the methods for all budgets $K\in\{0.5,1.0,\dots,7.0\}$.  After the methods finished, to compare their solutions, we separately evaluated the outputted solutions using $1000$ Monte-Carlo simulations each. The experiments were run out on a computer with an Intel Core i5-5200U CPU with 4 cores running at 2.2 GHz, and 8 GB of memory. We used Python for our implementation, along with the CyNetDiff library \cite{robson2024cynetdiff} for faster execution of the Monte-Carlo diffusion simulations. In addition, our implementation used an optimization similar to that of the CELF algorithm \cite{celf-07} for improved runtime when selecting the largest in the inner loop.

%The source codes of RIS-based greedy algorithm \cite{guo2020influence} and LIM-Greedy \cite{chen2020scalable} provided by their authors are written in C++.

\begin{figure*}[t!]
\centering
\begin{minipage}{.32\textwidth}
  \centering
  \resizebox{\textwidth}{!}{\input{figures/ic-wc/sim_facebook_plot_20}}
  \subcaption{Facebook network} \label{fig:influence_vs_budget_facebook}
\end{minipage}\hspace{.2cm}
\begin{minipage}{.32\textwidth}
  \centering
  \resizebox{\textwidth}{!}{\input{figures/ic-wc/sim_wikipedia_plot_20}}
  \subcaption{Wikipedia network} \label{fig:influence_vs_budget_wikipedia}
\end{minipage}\hspace{.2cm}
\begin{minipage}{.32\textwidth}
  \centering
  \resizebox{\textwidth}{!}{\input{figures/ic-wc/sim_deezer_plot_20}}
  \subcaption{Deezer network} \label{fig:influence_vs_budget_deezer}
\end{minipage}
\caption{Influence vs. Budget ($K$) for different networks.} \label{fig:influence_vs_budget}
\end{figure*}
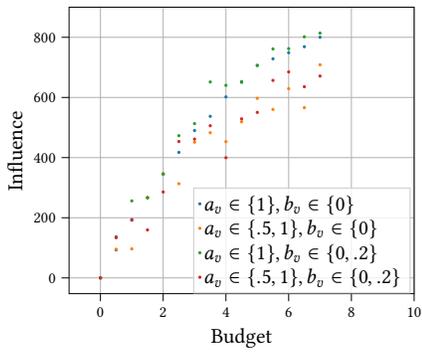
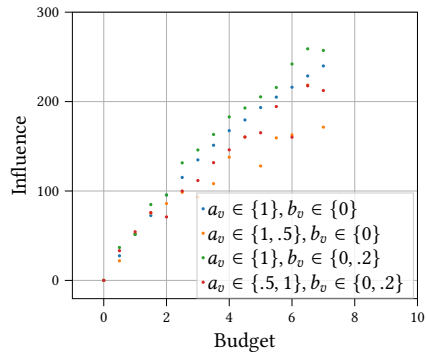
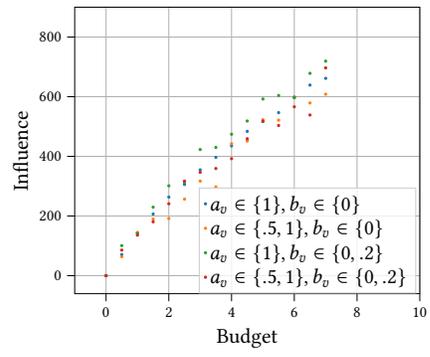

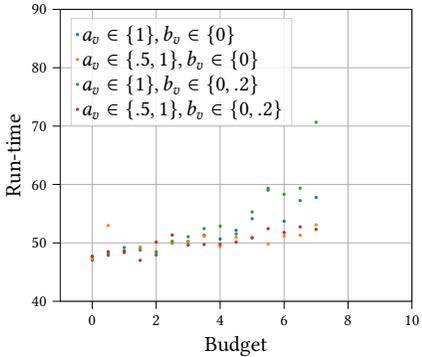
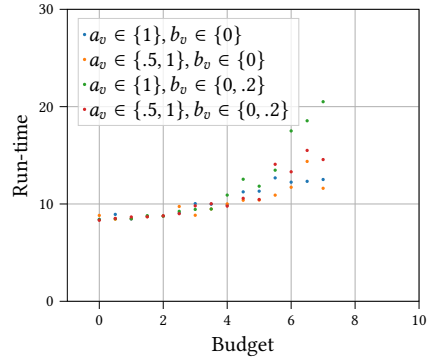
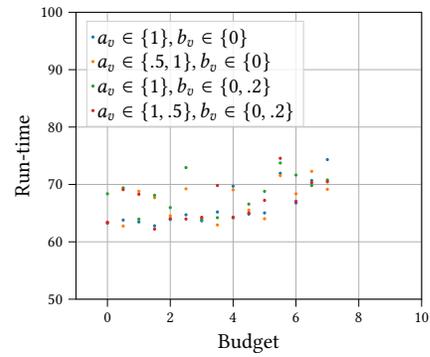
\begin{figure*}[h!]
\centering
\begin{minipage}{.32\textwidth}
  \centering
  \resizebox{\textwidth}{!}{\input{figures/ic-wc/sim_facebook_plot_20_ln_runtimes_avg}}
  \subcaption{Facebook network} \label{fig:runtime_vs_budget_facebook}
\end{minipage}\hspace{.25cm}
\begin{minipage}{.32\textwidth}
  \centering
  \resizebox{\textwidth}{!}{\input{figures/ic-wc/sim_wikipedia_plot_20_ln_runtimes_avg}}
  \subcaption{Wikipedia network} \label{fig:runtime_vs_budget_wikipedia}
\end{minipage}\hspace{.2cm}
\begin{minipage}{.32\textwidth}
  \centering
  \resizebox{\textwidth}{!}{\input{figures/ic-wc/sim_deezer_plot_20_ln_runtimes_avg}}
  \subcaption{Deezer network} \label{fig:runtime_vs_budget_deezer}
\end{minipage}
\caption{Runtime (in seconds) vs. Budget ($K$) for different networks. %plotted on a log scale.
} \label{fig:runtime_vs_budget}
\end{figure*}

\textbf{Results.}
\Cref{fig:influence_vs_budget} provides the influence (on $y$-axis) using different functions for different values of the budget (on $x$-axis) for different networks. The influence values in the plots are the expected numbers calculated using 1000 Monte-Carlo simulations. \Cref{fig:runtime_vs_budget} depicts the runtime in seconds ($y$-axis) using different functions for different values of the budget ($x$-axis), for different networks. The runtimes are not averaged, but relatively consistent across different budgets.

\textbf{Discussion.} The runtimes shown in \Cref{fig:runtime_vs_budget} demonstrate a consistently good performance of the proposed algorithm across different budgets and coefficient values $a_v$ and $b_v$ for $f_v(\cdot)$. Specifically, the scale used on each of these plots only covered a small range, showing similar performance (in terms of runtime) over the different parameters used in the experiments across different budgets. This consistency in runtime across budgets is most pronounced for the Deezer network (refer to \Cref{fig:influence_vs_budget_deezer}). Furthermore, as can be seen in \Cref{tab:basic_info}, the Deezer network has substantially more nodes than the Wikipedia and Facebook networks while achieving similar runtime with our algorithm. This demonstrates the scalability of our algorithm even for large networks.

Turning our attention to \Cref{fig:influence_vs_budget}, the relative similarity of final influence values across different coefficients is an indication of our algorithm's robustness. This shows our algorithm can handle more realistic inputs while still achieving good final output values. As a point of comparison, choosing the coefficients $a_v = 1, b_v = 0$ in the proposed algorithm recovers the behavior of the MLE-Greedy algorithm \cite{UmrawalAQ2023}. Thus, our results show that our algorithm achieves comparable performance to a known baseline while providing increased flexibility when given more diverse input data.

%% file: figures/ic-wc/sim_facebook_plot_20.tex
% This file was created by tikzplotlib v0.9.2.
\begin{tikzpicture}

\definecolor{color0}{rgb}{0.12156862745098,0.466666666666667,0.705882352941177}
\definecolor{color1}{rgb}{1,0.498039215686275,0.0549019607843137}
\definecolor{color2}{rgb}{0.172549019607843,0.627450980392157,0.172549019607843}
\definecolor{color3}{rgb}{0.83921568627451,0.152941176470588,0.156862745098039}
\definecolor{color4}{rgb}{0.580392156862745,0.403921568627451,0.741176470588235}

\begin{axis}[
legend cell align={left},
legend style={fill opacity=0.8, draw opacity=1, text opacity=1, at={(0.36,0.37)}, anchor=north west, draw=white!80!black},
tick align=outside,
tick pos=left,
x grid style={white!69.0196078431373!black},
xlabel={\LARGE Budget},
xmajorgrids,
xmin=-1, xmax=10,
xtick style={color=black},
y grid style={white!69.0196078431373!black},
ylabel={\LARGE Influence},
ymajorgrids,
ymin=-52.401, ymax=900,
ytick style={color=black}
]

\addplot [thick, only marks, color0, mark=*, mark size=.5, mark options={solid}]
table {%
0.0 0.000
0.5 93.092
1.0 193.995
1.5 266.267
2.0 344.296
2.5 417.355
3.0 490.069
3.5 537.326
4.0 601.849
4.5 650.287
5.0 706.742
5.5 728.443
6.0 748.763
6.5 768.776
7.0 800.184
};
\addlegendentry{\LARGE $a_v \in \{1\}, b_v \in \{0\}$}
\addplot [thick, only marks, color1, mark=*, mark size=.5, mark options={solid}]
table {%
0.0 0.000   
0.5 95.444  
1.0 96.597  
1.5 267.558  
2.0 345.792   
2.5 313.167   
3.0 451.372   
3.5 482.796   
4.0 453.056   
4.5 519.057   
5.0 597.063   
5.5 560.228   
6.0 629.170 
6.5 566.061
7.0 708.598
};
\addlegendentry{\LARGE $a_v \in \{.5,1\}, b_v \in \{0\}$}
\addplot [thick, only marks, color2, mark=*, mark size=0.5, mark options={solid}]
table {%
0.0 0.000
0.5 136.967
1.0 256.135
1.5 266.815
2.0 346.374
2.5 472.871
3.0 512.965
3.5 651.427
4.0 640.255
4.5 652.831
5.0 706.064
5.5 761.320
6.0 762.233
6.5 801.670
7.0 813.853
};
\addlegendentry{\LARGE $a_v \in \{1\}, b_v \in \{0,.2\}$}
\addplot [thick, only marks, color3, mark=*, mark size=0.5, mark options={solid}]
table {%
0.0 0.000
0.5 133.692
1.0 192.011
1.5 159.590
2.0 285.696
2.5 453.520
3.0 461.448
3.5 505.986
4.0 399.652
4.5 528.934
5.0 550.414
5.5 656.676
6.0 684.873
6.5 635.726
7.0 671.143
};
\addlegendentry{\LARGE $a_v \in \{.5,1\}, b_v \in \{0,.2\}$}
\end{axis}

\end{tikzpicture}

%% file: figures/ic-wc/sim_wikipedia_plot_20.tex
% This file was created by tikzplotlib v0.9.2.
\begin{tikzpicture}

\definecolor{color0}{rgb}{0.12156862745098,0.466666666666667,0.705882352941177}
\definecolor{color1}{rgb}{1,0.498039215686275,0.0549019607843137}
\definecolor{color2}{rgb}{0.172549019607843,0.627450980392157,0.172549019607843}
\definecolor{color3}{rgb}{0.83921568627451,0.152941176470588,0.156862745098039}
\definecolor{color4}{rgb}{0.580392156862745,0.403921568627451,0.741176470588235}

\begin{axis}[
legend cell align={left},
legend style={fill opacity=0.8, draw opacity=1, text opacity=1, at={(0.36,0.37)}, anchor=north west, draw=white!80!black},
tick align=outside,
tick pos=left,
x grid style={white!69.0196078431373!black},
xlabel={\LARGE Budget},
xmajorgrids,
xmin=-1, xmax=10,
xtick style={color=black},
y grid style={white!69.0196078431373!black},
ylabel={\LARGE Influence},
ymajorgrids,
ymin=-20.401, ymax=300,
ytick style={color=black}
]

\addplot [thick, only marks, color0, mark=*, mark size=.5, mark options={solid}]
table {%
0.0 0.000
0.5 27.533
1.0 51.344
1.5 72.583
2.0 95.537
2.5 115.157
3.0 134.814
3.5 151.154
4.0 167.482
4.5 179.475
5.0 193.325
5.5 205.000
6.0 216.100
6.5 228.662
7.0 239.855
};
\addlegendentry{\LARGE $a_v \in \{1\}, b_v \in \{0\}$}
\addplot [thick, only marks, color1, mark=*, mark size=.5, mark options={solid}]
table {%
0.0 0.000
0.5 21.894
1.0 52.144
1.5 75.250
2.0 86.039
2.5 98.607
3.0 93.178
3.5 108.240
4.0 137.802
4.5 160.240
5.0 127.961
5.5 159.458
6.0 162.719
6.5 218.493
7.0 171.417
};
\addlegendentry{\LARGE $a_v \in \{1,.5\}, b_v \in \{0\}$}
\addplot [thick, only marks, color2, mark=*, mark size=0.5, mark options={solid}]
table {%
0.0 0.000
0.5 36.837
1.0 51.650
1.5 84.852
2.0 95.499
2.5 131.466
3.0 145.891
3.5 163.237
4.0 182.850
4.5 192.848
5.0 205.272
5.5 215.783
6.0 242.028
6.5 259.026
7.0 257.226
};
\addlegendentry{\LARGE $a_v \in \{1\}, b_v \in \{0,.2\}$}
\addplot [thick, only marks, color3, mark=*, mark size=0.5, mark options={solid}]
table {%
0.0 0.000
0.5 33.221
1.0 54.397
1.5 75.927
2.0 70.996
2.5 99.971
3.0 111.751
3.5 131.716
4.0 146.054
4.5 160.453
5.0 165.140
5.5 194.486
6.0 160.186
6.5 217.679
7.0 212.363
};
\addlegendentry{\LARGE $a_v \in \{.5,1\}, b_v \in \{0,.2\}$}
\end{axis}

\end{tikzpicture}

%% file: figures/ic-wc/sim_deezer_plot_20.tex
% This file was created by tikzplotlib v0.9.2.
\begin{tikzpicture}

\definecolor{color0}{rgb}{0.12156862745098,0.466666666666667,0.705882352941177}
\definecolor{color1}{rgb}{1,0.498039215686275,0.0549019607843137}
\definecolor{color2}{rgb}{0.172549019607843,0.627450980392157,0.172549019607843}
\definecolor{color3}{rgb}{0.83921568627451,0.152941176470588,0.156862745098039}
\definecolor{color4}{rgb}{0.580392156862745,0.403921568627451,0.741176470588235}

\begin{axis}[
legend cell align={left},
legend style={fill opacity=0.8, draw opacity=1, text opacity=1, at={(0.36,0.37)}, anchor=north west, draw=white!80!black},
tick align=outside,
tick pos=left,
x grid style={white!69.0196078431373!black},
xlabel={\LARGE Budget},
xmajorgrids,
xmin=-1, xmax=10,
xtick style={color=black},
y grid style={white!69.0196078431373!black},
ylabel={\LARGE Influence},
ymajorgrids,
ymin=-60.401, ymax=900,
ytick style={color=black}
]
\addplot [thick, only marks, color0, mark=*, mark size=.5, mark options={solid}]
table {%
0.0 0.000
0.5 70.630
1.0 142.394
1.5 206.696
2.0 263.394
2.5 305.899
3.0 354.799
3.5 396.605
4.0 435.155
4.5 483.858
5.0 516.757
5.5 546.635
6.0 598.891
6.5 639.010
7.0 661.530
};
\addlegendentry{\LARGE $a_v \in \{1\}, b_v \in \{0\}$}
\addplot [thick, only marks, color1, mark=*, mark size=.5, mark options={solid}]
table {%
0.0 0.000   
0.5 63.591   
1.0 144.825   
1.5 189.711   
2.0 191.697   
2.5 256.629   
3.0 317.087   
3.5 297.553   
4.0 442.111   
4.5 451.388   
5.0 522.695   
5.5 521.387   
6.0 596.620  
6.5 578.958   
7.0 608.249 
};
\addlegendentry{\LARGE $a_v \in \{.5,1\}, b_v \in \{0\}$}
\addplot [thick, only marks, color2, mark=*, mark size=0.5, mark options={solid}]
table {%
0.0 0.000   
0.5 100.734   
1.0 142.003   
1.5 229.629   
2.0 301.204   
2.5 312.743   
3.0 422.807   
3.5 429.962   
4.0 474.197   
4.5 518.473   
5.0 592.267   
5.5 604.149   
6.0 596.651
6.5 678.389
7.0 719.300
};
\addlegendentry{\LARGE $a_v \in \{1\}, b_v \in \{0,.2\}$}
\addplot [thick, only marks, color3, mark=*, mark size=0.5, mark options={solid}]
table {%
0.0 0.000
0.5 85.808
1.0 135.887
1.5 180.575
2.0 241.043
2.5 317.175
3.0 346.437
3.5 359.260
4.0 392.298
4.5 459.116
5.0 517.989
5.5 503.506
6.0 566.210
6.5 538.667
7.0 696.590
};
\addlegendentry{\LARGE $a_v \in \{.5,1\}, b_v \in \{0,.2\}$}
\end{axis}

\end{tikzpicture}

%% file: figures/ic-wc/sim_facebook_plot_20_ln_runtimes_avg.tex
% This file was created by tikzplotlib v0.9.8.
\begin{tikzpicture}

\definecolor{color0}{rgb}{0.12156862745098,0.466666666666667,0.705882352941177}
\definecolor{color1}{rgb}{1,0.498039215686275,0.0549019607843137}
\definecolor{color2}{rgb}{0.172549019607843,0.627450980392157,0.172549019607843}
\definecolor{color3}{rgb}{0.83921568627451,0.152941176470588,0.156862745098039}
\definecolor{color4}{rgb}{0.580392156862745,0.403921568627451,0.741176470588235}

\begin{axis}[
legend cell align={left},
legend style={
  fill opacity=0.8,
  draw opacity=1,
  text opacity=1,
  at={(0.03,0.97)},
  anchor=north west,
  draw=white!80!black
},
log basis y={10},
tick align=outside,
tick pos=left,
x grid style={white!69.0196078431373!black},
xlabel={\LARGE Budget},
xmajorgrids,
xmin=-1, xmax=10,
xtick style={color=black},
y grid style={white!69.0196078431373!black},
ylabel={\LARGE Run-time},
ymajorgrids,
ymin=40, ymax=90,
ytick style={color=black}
]
\addplot [thick, only marks, color0, mark=*, mark size=.5, mark options={solid}]
table {%
0.0   47.054570
0.5   47.878089
1.0   48.681316
1.5   49.199589
2.0   47.930344
2.5   49.991845
3.0   50.250049
3.5   51.315582
4.0   50.673838
4.5   52.156815
5.0   54.142511
5.5   59.337611
6.0   53.714698
6.5   57.247753
7.0   57.774661
};
\addlegendentry{\LARGE $a_v \in \{1\}, b_v \in \{0\}$}
\addplot [thick, only marks, color1, mark=*, mark size=.5, mark options={solid}]
table {%
0.0   47.336713
0.5   52.966461
1.0   48.300082
1.5   49.288918
2.0   48.526404
2.5   49.915653
3.0   50.116690
3.5   51.198036
4.0   49.371462
4.5   50.946525
5.0   50.924490
5.5   49.826064
6.0   51.170493
6.5   51.333059
7.0   53.087676
};
\addlegendentry{\LARGE $a_v \in \{.5,1\}, b_v \in \{0\}$}
\addplot [thick, only marks, color2, mark=*, mark size=0.5, mark options={solid}]
table {%
0.0   47.666353
0.5   48.227314
1.0   49.221109
1.5   48.759484
2.0   48.388971
2.5   50.271885
3.0   51.057278
3.5   52.444663
4.0   52.873005
4.5   51.543514
5.0   55.300309
5.5   59.014914
6.0   58.312436
6.5   59.368466
7.0   70.659881
};
\addlegendentry{\LARGE $a_v \in \{1\}, b_v \in \{0,.2\}$}
\addplot [thick, only marks, color3, mark=*, mark size=0.5, mark options={solid}]
table {%
0.0   47.709761
0.5   48.494321
1.0   48.440297
1.5   47.012870
2.0   50.147527
2.5   51.349247
3.0   49.656516
3.5   49.729057
4.0   49.756237
4.5   50.133785
5.0   50.863954
5.5   52.444692
6.0   51.788360
6.5   52.742997
7.0   52.316993
};
\addlegendentry{\LARGE $a_v \in \{.5,1\}, b_v \in \{0,.2\}$}
\end{axis}

\end{tikzpicture}

%% file: figures/ic-wc/sim_wikipedia_plot_20_ln_runtimes_avg.tex
% This file was created by tikzplotlib v0.9.8.
\begin{tikzpicture}

\definecolor{color0}{rgb}{0.12156862745098,0.466666666666667,0.705882352941177}
\definecolor{color1}{rgb}{1,0.498039215686275,0.0549019607843137}
\definecolor{color2}{rgb}{0.172549019607843,0.627450980392157,0.172549019607843}
\definecolor{color3}{rgb}{0.83921568627451,0.152941176470588,0.156862745098039}
\definecolor{color4}{rgb}{0.580392156862745,0.403921568627451,0.741176470588235}

\begin{axis}[
legend cell align={left},
legend style={
  fill opacity=0.8,
  draw opacity=1,
  text opacity=1,
  at={(0.03,0.97)},
  anchor=north west,
  draw=white!80!black
},
log basis y={10},
tick align=outside,
tick pos=left,
x grid style={white!69.0196078431373!black},
xlabel={\LARGE Budget},
xmajorgrids,
xmin=-1, xmax=10,
xtick style={color=black},
y grid style={white!69.0196078431373!black},
ylabel={\LARGE Run-time},
ymajorgrids,
ymin=0, ymax=30,
ytick style={color=black}
]
\addplot [thick, only marks, color0, mark=*, mark size=.5, mark options={solid}]
table {%
0.0    8.405542
0.5    8.930355
1.0    8.456910
1.5    8.707745
2.0    8.747031
2.5    9.067738
3.0   10.035256
3.5    9.981037
4.0    9.875415
4.5   11.244788
5.0   11.315327
5.5   12.679956
6.0   12.239418
6.5   12.326786
7.0   12.514295
};
\addlegendentry{\LARGE $a_v \in \{1\}, b_v \in \{0\}$}
\addplot [thick, only marks, color1, mark=*, mark size=.5, mark options={solid}]
table {%
0.0    8.825514
0.5    8.432885
1.0    8.489856
1.5    8.694812
2.0    8.759588
2.5    9.744255
3.0    8.841070
3.5    9.489387
4.0   10.007888
4.5   10.382982
5.0   10.403745
5.5   10.906649
6.0   11.716111
6.5   14.373694
7.0   11.611880
};
\addlegendentry{\LARGE $a_v \in \{.5, 1\}, b_v \in \{0\}$}
\addplot [thick, only marks, color2, mark=*, mark size=0.5, mark options={solid}]
table {%
0.0    8.394794
0.5    8.488534
1.0    8.488526
1.5    8.799099
2.0    8.757524
2.5    9.244291
3.0    9.442190
3.5    9.479621
4.0   10.913176
4.5   12.541381
5.0   11.825968
5.5   13.472525
6.0   17.505825
6.5   18.548547
7.0   20.506877
};
\addlegendentry{\LARGE $a_v \in \{1\}, b_v \in \{0,.2\}$}
\addplot [thick, only marks, color3, mark=*, mark size=0.5, mark options={solid}]
table {%
0.0    8.325471
0.5    8.517986
1.0    8.666395
1.5    8.686916
2.0    8.786479
2.5    9.002386
3.0    9.806827
3.5   10.025653
4.0    9.780839
4.5   10.581501
5.0   10.463497
5.5   14.077605
6.0   13.306242
6.5   15.503675
7.0   14.563195
};
\addlegendentry{\LARGE $a_v \in \{.5,1\}, b_v \in \{0,.2\}$}
\end{axis}

\end{tikzpicture}

%% file: figures/ic-wc/sim_deezer_plot_20_ln_runtimes_avg.tex
% This file was created by tikzplotlib v0.9.8.
\begin{tikzpicture}

\definecolor{color0}{rgb}{0.12156862745098,0.466666666666667,0.705882352941177}
\definecolor{color1}{rgb}{1,0.498039215686275,0.0549019607843137}
\definecolor{color2}{rgb}{0.172549019607843,0.627450980392157,0.172549019607843}
\definecolor{color3}{rgb}{0.83921568627451,0.152941176470588,0.156862745098039}
\definecolor{color4}{rgb}{0.580392156862745,0.403921568627451,0.741176470588235}

\begin{axis}[
legend cell align={left},
legend style={
  fill opacity=0.8,
  draw opacity=1,
  text opacity=1,
  at={(0.03,0.97)},
  anchor=north west,
  draw=white!80!black
},
log basis y={10},
tick align=outside,
tick pos=left,
x grid style={white!69.0196078431373!black},
xlabel={\LARGE Budget},
xmajorgrids,
xmin=-1, xmax=10,
xtick style={color=black},
y grid style={white!69.0196078431373!black},
ylabel={\LARGE Run-time},
ymajorgrids,
ymin=50, ymax=100,
ytick style={color=black}
]
\addplot [thick, only marks, color0, mark=*, mark size=.5, mark options={solid}]
table {%
0.0   63.254205
0.5   63.790675
1.0   63.461924
1.5   62.784659
2.0   63.895771
2.5   64.711532
3.0   63.656964
3.5   65.195508
4.0   69.696210
4.5   64.827621
5.0   65.036040
5.5   71.937986
6.0   66.766276
6.5   70.682599
7.0   74.329576
};
\addlegendentry{\LARGE $a_v \in \{1\}, b_v \in \{0\}$}
\addplot [thick, only marks, color1, mark=*, mark size=.5, mark options={solid}]
table {%
0.0   63.447013
0.5   62.748536
1.0   68.806477
1.5   67.690603
2.0   64.521267
2.5   69.241434
3.0   64.114415
3.5   62.931424
4.0   69.044469
4.5   65.523077
5.0   64.021747
5.5   71.544206
6.0   68.375939
6.5   72.290093
7.0   69.153528
};
\addlegendentry{\LARGE $a_v \in \{.5,1\}, b_v \in \{0\}$}
\addplot [thick, only marks, color2, mark=*, mark size=0.5, mark options={solid}]
table {%
0.0   68.372162
0.5   69.399346
1.0   63.950276
1.5   68.111505
2.0   65.968659
2.5   72.935726
3.0   63.879235
3.5   64.200661
4.0   64.161983
4.5   66.569758
5.0   68.800764
5.5   73.739130
6.0   71.634069
6.5   69.803899
7.0   70.773218
};
\addlegendentry{\LARGE $a_v \in \{1\}, b_v \in \{0,.2\}$}
\addplot [thick, only marks, color3, mark=*, mark size=0.5, mark options={solid}]
table {%
0.0   63.380199
0.5   69.091620
1.0   68.291425
1.5   62.214077
2.0   64.051993
2.5   63.971418
3.0   64.269888
3.5   69.814956
4.0   64.290657
4.5   65.039678
5.0   67.225920
5.5   74.555295
6.0   67.060689
6.5   70.300362
7.0   70.457565
};
\addlegendentry{\LARGE $a_v \in \{1,.5\}, b_v \in \{0,.2\}$}
\end{axis}

\end{tikzpicture}

%% file: sections/conclusion.tex
\section{Conclusion}
\label{sec:conclusion}
% For the problem of influence maximization under marketing strategies involving an affine relation (potentially varying across users) between the adoption likelihood and the incentive received by a user, we provided an efficient $(1-1/e)$-approximation algorithm and demonstrated its performance using experimental evaluations. In the future, we want to explore a convex or arbitrary non-linear relation between the adoption likelihood and the incentive received by a user and extend our work to an online setting \cite{agarwal2022stochastic,nie2022explore}. 

We studied the problem of influence maximization over social networks under general marketing strategies. We specifically focused on strategies involving an affine relation (potentially varying across users) between the adoption likelihood and the incentive received by a user. We proposed an efficient $(1-1/e)$-approximation algorithm and demonstrated its performance using experimental evaluations. In the future, we are interested in exploring even more general marketing strategies. For instance, a convex relation and an arbitrary non-linear relation between the adoption likelihood and the incentive received by a user. We are also interested in extending our work to an online setting \cite{agarwal2022stochastic,nie2022explore}.